\documentclass[english]{sbrt}
\usepackage[english]{babel}
\usepackage[utf8]{inputenc}

\usepackage[pdftex]{color,graphicx}
\usepackage{wrapfig}
\usepackage{amsfonts}
\usepackage{amssymb}
\usepackage{physics}

\newtheorem{thm}{Theorem} [section] 
\newtheorem{lem}[thm]{Lemma} 
\newtheorem{prop}[thm]{Proposition} 

\newtheorem{defn}[thm]{Definition} 
\newtheorem{ex}[thm]{Example} 

\begin{document}

    \title{A condition for the zero-error capacity of quantum channels}
    
    \author{Marciel M. Oliveira, Francisco M. de Assis and  Micael A. Dias
    \thanks{Marciel M. de Oliveira; Francisco M. de Assis and  Micael A. Dias are with the Department of  Electrical  Engineering,  Federal  University  of Campina Grande, Paraíba,  Brazil. E-mails: marciel.oliveira@ee.ufcg.edu.br, fmarcos@dee.ufcg.edu.br, micael.souza@ee.ufcg.edu.br. This work was partially funded by the CNPq Project (305918/2019-2) and CAPES/PROEXT.}%
    }
    
    \maketitle
    
    \markboth{XLI BRAZILIAN SYMPOSIUM ON TELECOMMUNICATIONS AND SIGNAL PROCESSING - SBrT 2023, OCTOBER 08--11, 2023, S\~{A}O JOSÉ DOS CAMPOS, SP}{}

    \begin{abstract}
    In this paper, we present a condition for the zero-error capacity of quantum channels. To achieve this result we first prove that the eigenvectors (or eigenstates) common to the Kraus operators representing the quantum channel are fixed points of the channel. From this fact and assuming that these Kraus operators have at least two eigenstates in common and also considering that every quantum channel has at least one fixed point, it is proved that the zero-error capacity of the quantum channel is positive. Moreover, this zero-error capacity condition is a lower bound for the zero-error capacity of the quantum channel. This zero-error capacity condition of quantum channels has a peculiar feature that it is easy to verify when one knows the Kraus operators representing the quantum channel.    
    \end{abstract}

    \section{Introduction}
    
        Information theory, in general, is understood as a field of study in which the central elements are the quantification, storage, and transmission of information. In the classical scenario, classical information theory has \textit{bits} as the basic elements, while the transmission, processing and storage of information are according to the laws of classical mechanics \cite{Cover}. In the quantum scenario, quantum information theory is seen as a new paradigm for information processing and information transmission through quantum channels, while taking into account the laws of quantum mechanics. Therefore, it means that information is represented not only in the form of \textit{bits}, but also in the form of \textit{quantum bits}. However, the approach to the concepts of information theory in the quantum scenario is carried out with a strategy analogous to the classical scenario \cite{Chuang,Wilde}.
        
        One of the main topics in classical and quantum information theory is the study of how much information can be transmitted reliably through a physical medium. The (ordinary) capacity of classical channels is a number that indicates the asymptotic rate at which information can be reliably transmitted through the channel. The properties of quantum mechanics allow the capacity of a quantum channel to be defined in several ways, each of them depending on the type of information to be transmitted (classical information or quantum states), physical characteristics used in the transition, such as entanglement, and the communication protocol used. For the transmission of classical information, there are two examples of quantum channel capacity definitions. 
        
        The capacity of a quantum channel is defined by Holevo-Schumacher-Westmoreland (HSW) and is defined as the maximum asymptotic rate at which classical information can be reliably transmitted using quantum encoding and decoding \cite{HSW,Holevo}. The other example of the capacity for quantum channels is the classical entanglement capacity $C_E$, which is defined as the supreme rate of transmission of classical information over a quantum channel when an infinite amount of entanglement is available between the transmitter and receiver \cite{Bennett}. 
        
        These capacities defined above admit a probability of error greater than zero, even if the best encoding scheme is used. Sometimes it is interesting to transmit information with an error rate error probability equal to zero \cite{Koner}. The zero-error capacity of quantum channels was defined by Medeiros and Assis \cite{Rex0} as the supreme rate at which classical information can be transmitted over a quantum channel with error probability exactly equal to zero. In this way, using a quantum block code, classical information is encoded into quantum states, and it is transmitted over a discrete memoryless channel (DMC). At reception, as part of the process, the quantum states are measured. In this case, the transmission is performed with an error probability of exactly zero. The zero-error capacity of quantum channels is a generalization of the zero-error capacity of DMC that was first defined by Shannon \cite{Shannon}. 
        
        Since the definition of zero-error capacity has been cited in the scientific community, recently Yamasaki and Murao \cite{ Yamasaki}, in a research involving quantum entanglement cost, mentions the pioneering definition.  In another paper, a study on the characteristic graph of quantum channels, Dereniowski and Jurkiewicz \cite{Dereniowski} suggest that more recent ones are in \cite{Rex1}.
        
        An important aspect in studying the zero-error capacity of quantum channels is to verify whether a quantum channel has a positive zero-error capacity or not. Three conditions are known in the literature to verify the condition of zero-error capacity of quantum channels, which are found in Medeiros and Rex \cite{Rex0,Rex2} and Gupta \textit{et al.} \cite{Ved}. In this paper, another condition will be presented to verify the zero-error capacity of quantum channels, which takes into account whether the Kraus operators representing the quantum channel have at least two common eigenstates. 
        
        This paper is organized in the following structure. Section II recalls the definition of the zero-error capacity of a quantum channel, represented by a completely positive trace-preserving linear map. In Section III, the condition for zero-error capacity of quantum channels is proved. Furthermore, some examples illustrating the condition and also an example for the non-validity of the reciprocal of the condition are presented. Finally, Section IV presents the conclusions and discussions of this work.
        
        \subsection{Notation and Conventions} 
            Dirac notation will be used to denote quantum states and operations on them \cite{Dirac}. The symbol $\mathbb{I}$ denotes the identity operator. Logarithms are used in base 2.
    
    \section{Zero-error capacity of quantum channels}
    
        In this section, the main definitions and results on the zero-error capacity of quantum channels are discussed, which are important to facilitate the reading and understanding of this paper. The discussions and notations used are based on \cite{Rex0,Rex1}.  
        
        Let $\mathcal{H}$ be a Hilbert space of dimension $d$. A quantum channel\footnote{ For more details on quantum channels, we refer to Section 3.3.2, \cite{Rex1}. } $\mathcal{E}$ defined on $\mathcal{H}$ is a completely positive and trace-preserving linear map\footnote{A discussion of completely positive and trace-preserving linear map we suggest reading Section 4.4, \cite{Wilde}.}  of density operators\footnote{An approach on density operators can be found in Section 2.4, \cite{Rex1}.} and can be described by a set $\{A_{i}\}$ of Kraus operators in $\mathcal{H}$ that satisfy the conditions $\sum_{i}^{\kappa}A_i^\dagger A_i=\mathbb{I}$ and $\kappa \leq d^2$.
        
        Let $\mathcal{X} \subset \mathcal{H}$ be the set of possible input states for the quantum channel $\mathcal{E}$. If $ \rho \in \mathcal{X}$, denote by 
        $\sigma = \mathcal{E}(\rho)$, the quantum state received when $\rho$ is transmitted over the quantum channel and can be written as
        \begin{equation} 
        \mathcal{E}(\rho)=\sum_{i=1}^{\kappa}A_i\rho A_i^\dagger.
        \end{equation}              
        
        The idea of a communication protocol over the quantum channel $\mathcal{E}$ associated with its zero-error capacity is summarized below. Define the finite subset $\mathcal{S}=\{\rho_1,...,\rho_\ell\} \subset \mathcal{X}$ and let $ \rho_i \in \mathcal{S}$.  The states $ \rho_i$ form the alphabet of a quantum zero-error code.  The set of code words of length $n$ is a superset of the sequences of $n$ tensor products of $\mathcal{S}$ states, denoted by $\mathcal{S}^{\otimes n}$. For $\rho_{i_j}$ the $i$-th codeword is given by $\overline{\rho}_i=\rho_{i_1}\otimes...\otimes \rho_{i_n}.$ Bob performs measurements using a POVM \textit{(Positive Operator-Valued Measurements)} with elements $\{M_j\}$, where $\sum_{j}^{}M_j= \mathbb{I}$. Then, $p(j|i)$ is the probability that Bob measures the outcome $j$ given that Alice sent state $\rho_i$. Thus,
        \begin{equation} 
        p(j|i)=\mbox{tr}[\sigma_iM_j]=\mbox{tr} [\mathcal{E}(\rho_i)M_j].
        \end{equation}
        
        A quantum $(m,n)$ zero-error code for $\mathcal{E}$ is composed of:
        \begin{itemize}
            \item[1)] A set of indices $\{1, . . . ,m\}$, where each index is associated with a classical message.
            \vspace{0.2cm}
            \item[2)] A coding function
                \begin{equation}
                    f_n:\{1,...,m\}\longrightarrow \mathcal{S}^{\otimes n} 
                \end{equation}
                which takes each codeword originating codewords $f_n(1)=\overline{\rho_1},...,f_n(m)=\overline{\rho_m}.$
                \vspace{0.2cm}
            \item[3)] A decoding function
                \begin{equation}
                g:\{1,...,k\}\longrightarrow \{1,...,m\} 
                \end{equation}
                which deterministically associates a message with one of the possible measurement $y \in \{1,...,k\}$ performed by the POVM $\{M_i\}_{i=1}^{k}$. Furthermore, the decoding function has the following property:
                \begin{equation}
                    \mbox{Pr}[g(\mathcal{E}(f_n(i)))\neq i]=0
                \end{equation}
                for all $i \in \{1,...,m\}.$
        \end{itemize}
        The rate of a $(m,n)$ code is given by 
        \begin{equation}
            R=\frac{1}{n} \log \ m\  \mbox{ bits/use.}
        \end{equation}
        
        Thus, the zero-error capacity of a quantum channel is defined as follows \cite{Rex0}.
        
        \begin{defn}
            The quantum zero-error capacity of a quantum channel $\mathcal{E}(\cdot)$, denoted by $C^{(0)}(\mathcal{E})$, is the supremum of the achievable rates with decoding error probability equal to zero,
            \begin{equation}
                C^{(0)}(\mathcal{E})=\sup_{\mathcal{S}}\sup_{n}\frac{1}{n}\log m
            \end{equation}
            where $m$ is the maximum number of classical messages the system can transmit without error when a zero-error quantum block code $(m,n)$ is used and the input alphabet is $\mathcal{S}.$
        \end{defn}
        \vspace{0.2cm}
        
        A fundamental property of quantum states is that not every pair of states can be perfectly distinguished. Two quantum states are distinguishable if and only if the Hilbert subspaces generated by the supports of these quantum states are orthogonal. Thus, given two quantum states $\ket{\rho_i},\ket{\rho_j} \in \mathcal{S}$ with $i\neq j$ , then $\ket{\rho_i}$ and $\ket{\rho_j} $ are said to be \textit{non-adjacent} (or distinguishable) at the output of the quantum channel $\mathcal{E}$, if $\mathcal{E}(\ket{\rho_i})$ and $\mathcal{E}(\ket{\rho_j})$ belong to orthogonal Hilbert subspaces. Otherwise, $\ket{\rho_i}$ and $\ket{\rho_j} $ are said to be \textit{adjacent} to (or indistinguishable) in the output of $\mathcal{E}$.
        
        Thus, the zero-error capacity conditions in \cite{Rex0}, \cite{Rex2} (see Proposition 1 and Proposition 3, respectively) show that a quantum channel $\mathcal{E}$ has positive zero-error capacity if and only if there are at least two non-adjacent quantum states. Equivalently this condition is proved in \cite{Ved} (See Lema 4.4.1) that the zero-error capacity of the channel $\mathcal{E}$ is positive if and only if $\ket{\rho_m}\bra{\rho_{m'}}$ is orthogonal to the subspace
        \begin{equation}
            S:=\mbox{span}\{A_i^\dagger A_j,\  i,j  \},
        \end{equation}
        where $\ket{\rho_m}$ and $\ket{\rho_{m'}}$ with $m\neq m'$ are input states in the channel and $A_i$ are the Kraus operators representing the quantum channel $\mathcal{E}$.
        
        There is a relationship between the zero-error capacity and the fixed point of the quantum channel $\mathcal{E}$. Schauder's fixed point theorem guarantees that every quantum channel has at least one quantum state $\rho$ that is a fixed point for the quantum channel $\mathcal{E}$, i.e., $\mathcal{E} (\rho)=\rho$. This relation is presented in the following proposition.
        \vspace{0.2cm}
        \begin{prop} [\cite{Rex0}] \label{fixed point}
        	Let $\mathcal{E}$ be a quantum channel with $N_f$ fixed points. Then the zero-error capacity of $\mathcal{E}$ is at least $log\ N_f.$
        \end{prop}
        
        The proof of this result can be found in \cite{Rex0}, Proposition 2.
        
        In the next section, a condition for the zero-error capacity of quantum channels is presented. This condition relates the Kraus operators that represent the channel and the number of common eigenstates that the operators have. The conclusion of the proof of the result uses Proposition \ref{fixed point} above. 
    
    \section{Condition for the zero-error capacity of quantum channels}
    
        To begin this section, some notation is introduced. Let $\mathcal{H}$ be a Hilbert space with dimension $d$. By $\mathcal{B}(\mathcal{H})$ we will denote the set of all linear continuous operators on $\mathcal{H}$. By $M_d(\mathbb{C})$ we denote the set of all complex square matrices of order $d$. Assumed that $\mathcal{H}\cong\mathbb{C}^d$ and $\mathcal{B}(\mathcal{H})\cong M_d(\mathbb{C}),$ then the quantum channel $\mathcal{E}$ can be viewed as a trace-preserving quantum superoperator $\mathcal{E}: \mathcal{B}(\mathcal{H})\longrightarrow \mathcal{B}(\mathcal{H}),$ or $\mathcal{E}: M_d(\mathbb{C})\longrightarrow M_d(\mathbb{C})$.
        
        In the following, it is proved that every common eigenstate of the Kraus operators representing the quantum channel is a fixed point of the channel.  From this result, it is possible to conclude that the zero-error capacity of the quantum channel is positive if the Kraus operators have at least two common eigenstates. 
        
        Before proving the following lemma we need to point out that for pure states $\ket{\psi}$ in the Hilbert space  $\mathcal{H}$ , we have that 
        \begin{equation} 
        \mathcal{E}(\ket{\psi})=\sum_{i=1}^{\kappa}A_i\op{\psi} A_i^\dagger.
        \end{equation}
        
        \begin{lem}\label{Lema do estado cumum}
            Let $\mathcal{E}: M_d(\mathbb{C})\longrightarrow M_d(\mathbb{C})$ be a quantum channel with Kraus operators $A_1,A_2,...,A_\kappa$. If $\ket{\psi}$ is a common eigenstate of the operators $A_i$ then it is a fixed point to $\mathcal{E}.$
        \end{lem}
        
        \begin{proof}
        We just need to show that $ \mathcal{E} (\ket{\psi}\bra{\psi})= \ket{\psi}\bra{\psi}$. For the sake of generality, consider $A_i\ket{\psi} =\lambda_i \ket{\psi}$, that is, it can be associated with different eigenvalues for different Kraus operators. Then, 
        \begin{align}
            \mathcal{E}(\op{\psi}) &\overset{(a)}{=} \sum_iE_i\op{\psi}E_i^\dagger\\
                                   &\overset{(b)}{=} \sum_i\lambda_i\ket{\psi}\lambda_i^*\bra{\psi}\\
                                   &\overset{(c)}{=} \op{\psi}\sum_i|\lambda_i|^2\\
                                   &\overset{(d)}{=} \op{\psi},
        \end{align}
        where in (a) we used the Kraus representation of the quantum channel, in (b) the fact that $\ket{\psi}$ is an eigenvector of every $A_i$ (by hypothesis) and (d) because $\mathcal{E}$ is trace-preserving.
        \end{proof}
        
        Now, define $N_\mathcal{E} = \{ \ket{\psi} \in \mathcal{S}\ :\  A_i\ket{\psi}=\lambda_i\ket{\psi} \}$ as the set eigenstates that are common for every Kraus operator for the quantum channel $\mathcal{E}$. Denote by $|N_\mathcal{E}|$ the cardinality of $N_\mathcal{E}.$
        
        \begin{thm} \label{Test cap. zero-error}
        	Let $N_f$ be the number of fixed points of a quantum channel $\mathcal{E}$. Then, $N_f \geq |N_\mathcal{E}|$ and $C^{(0)}(\mathcal{E}) \geq \mbox{log}\  |N_\mathcal{E}|.$
        \end{thm}
        
        \begin{proof}
            The lemma \ref{Lema do estado cumum} showed that the states in $N_\mathcal{E}$ are fixed points of the channel. Since a fixed point does not necessarily need to be a common eigenstate for each Kraus operator, the inequality $N_f \geq |N_\mathcal{E}|$ is valid. For the zero-error capacity lower bound, one can construct a trivial quantum block code by encoding classical information in states in $N_\mathcal{E}$ and then it is possible to transmit at least $\mbox{log}\  |N_\mathcal{E}|$ bits through the quantum channel without error. So, we have $C^{(0)}(\mathcal{E}) \geq \mbox{log}\  |N_\mathcal{E}|.$
        \end{proof}
        
        Note that Theorem \ref{Test cap. zero-error} 
        also functions as a bound on the zero-error capacity of a quantum channel. Therefore, by this condition, the zero-error capacity of a quantum channel is bounded below by $\log|N_\mathcal{E}|$.
        
        The reciprocal of Theorem \ref{Test cap. zero-error} is not valid, that is, a quantum channel $\mathcal{E}$ can have a positive zero-error capacity and its Kraus operators do not have a common eigenstate. The following example shows a quantum channel with positive zero-error capacity, but the Kraus operators have no common eigenstate.
        
        \begin{ex} Let $\mathcal{E}$ be the quantum channel represented by the Kraus operators $A_1$, $A_2$ and $A_3$ given by:
            \begin{align*}
                A_1 &= \mqty(0.5 & 0 & 0 & 0 & \frac{\sqrt{49902}}{620} \\
        	               0.5 & -0.5 &  0 & 0 & 0 \\
        	               0 & 0.5 &  -0.5 & 0 & 0\\
        	               0 & 0 &  0.5 & -\frac{\sqrt{457}}{50}  & \frac{\sqrt{457}}{50} \\
        	               0 & 0 &  0 & -0.62 & \frac{289}{1550}),\\
                A_2&=\mqty(0.5 & 0 & 0 & 0 & \frac{\sqrt{49902}}{620} \\
                	     0.5 & 0.5 &  0 & 0 & 0 \\
                	   0 & 0.5 &  0.5 & 0 & 0\\
                          0 & 0 &  0.5 & \frac{\sqrt{457}}{50}  & -\frac{\sqrt{457}}{50} \\
                	   0 & 0 &  0 & 0.5 & 0.5),\\
                A_3&= \mqty(0 & 0 & 0 & 0 & 0 \\
        	              0 & 0 &  0 & 0 & 0 \\
        	              0 & 0 &  0 & 0 & 0\\
        	              0 & 0 &  0 & 0  & 0 \\
        	              0 & 0 &  0 & 0 & 0.3).
            \end{align*}
            The Kraus operators $A_1$, $A_2$ and $A_3$ have no common eigenstate but the channel has positive zero-error capacity (see Example 5.4, \cite{Rex2}).
        \end{ex}
        
        The following are some examples illustrating the zero-error capacity condition of a quantum channel.
        
        \begin{ex} The quantum channel $\mathcal{E}$ with the Kraus operators $A_1$ and $A_2$, given by
            \begin{align*}
                A_1 &= \mqty(\frac{3}{10} & 0 & -\frac{3}{10}\\
                            0 & 0 &  0\\
                            -\frac{3}{10} & 0 & \frac{3}{10})\\
                A_2 &= \mqty(-\frac{1}{10} & 0 & -\frac{9}{10}\\
                            0 & 0 &  0\\
                            -\frac{9}{10} & 0 & -\frac{1}{10}),
            \end{align*}
            the three vectors $\ket{\psi}=(1,0,1)$, $\ket{\phi}=(0,1,0)$ and $\ket{\varphi}=(-1,0,1)$ are eigenstate common to $A_1$ and $A_2$. According to the Lemma \ref{Lema do estado cumum} these three eigenstates common to $A_1$ and $A_2$ are fixed points of the quantum channel. Moreover, by the condition presented in Theorem \ref{Test cap. zero-error}, it follows that $C^{(0)}(\mathcal{E})\geq log\ 3.$ 
        \end{ex}	
        	
        \begin{ex} Assume that $p \in (0,1)$ and consider the quantum channel $\mathcal{E}$ with the Kraus operators $A_1$ and $A_2$ below:
            \begin{align*}
                A_1 &= \mqty(1 & 0 & 0 & 0\\
                            0 & \frac{1}{2} &  \frac{\sqrt{3}}{2} & 0 \\
                            0 & \frac{\sqrt{3}}{2} &  -\frac{1}{2} & 0 \\
                            0 & 0 &  0 & 1)\sqrt{p},\\
                A_2 &= \mqty(1 & 0 & 0 & 0\\
                            0 & \frac{1}{2} & - \frac{\sqrt{3}}{2} & 0 \\
                            0 & -\frac{\sqrt{3}}{2} &  -\frac{1}{2} & 0 \\
                            0 & 0 &  0 & 1 )\sqrt{1-p},
            \end{align*}
            the two vectors $\ket{\psi}=(1,0,0,0)$ and $\ket{\varphi}=(0,0,0,1)$ are eigenstate common to $A_1$ and $A_2$. The Lemma \ref{Lema do estado cumum} proves that these two eigenstates common to $A_1$ and $A_2$ are fixed points of the quantum channel.  Thus, by the condition presented in Theorem \ref{Test cap. zero-error}, it follows that $C^{(0)}(\mathcal{E})\geq log\ 2 =1.$
        \end{ex}
        	 
         \begin{ex}
         Let be the quantum channel $\mathcal{E}$ represented by the Kraus operators $A_1$, $A_2$ and $A_3$ given by:
        \begin{align*}
            A_1&= \mqty(0 & 0 & 0 & 0 & 0.5 \\
                     0 & 0.5 &  0 & 0 & 0 \\
                     0 & 0&  0.5 & 0 & 0\\
                     0 & 0 &  0 & 0.5  & 0 \\
                     0 & 0 &  0 & 0 & 0.5)\\
            A_2&= \mqty(0 & 0 & 0 & 0 & -0.5 \\
                     0 & -0.5 &  0 & 0 & 0 \\
                     0 & 0&  -0.5 & 0 & 0\\
                     0 & 0 &  0 & -0.5  & 0 \\
                     0 & 0 &  0 & 0 & -0.5)\\
            A_3&= \mqty(1 & 0 & 0 & 0 & 0 \\
                     0 & \sqrt{0.5} &  0 & 0 & 0 \\
                     0 & 0&  \sqrt{0.5} & 0 & 0\\
                     0 & 0 &  0 & \sqrt{0.5} & 0 \\
                     0 & 0 &  0 & 0 & 0)
        \end{align*}
        The Kraus operators $A_1$, $A_2$ and $A_3$ have four common  eigenstate which are $\ket{\psi}=(1,0,0,0,0) $, $\ket{\phi}=(0,1,0,0,0) $, $\ket{\varphi}=(0,0,1,0,0) $ and $\ket{\mu}=(0,0,0,1,0)$, these eigenstates are fixed points of the quantum channel by the Lemma \ref{Lema do estado cumum}. By the zero-error capacity condition in Theorem \ref{Test cap. zero-error} we have $C^{(0)}(\mathcal{E})\geq log\ 4 = 2.$
        \end{ex}
        
        In these examples we present an illustration of the zero-error capacity condition of a quantum channel theorem. For this purpose, it was verified that the Kraus operators representing the quantum channel have eigenstates in common and they are fixed points of the channel, concluding that the number of eigenstates common to the Kraus operators is a lower bound on the zero-error capacity of the quantum channel.
    
    \section{Conclusions}
    
        In this paper, a zero-error capacity condition of quantum channels was presented. The condition shows that if the Kraus operators of the quantum channel have at least two common eigenstates, the channel has non-trivial zero-error capacity. A particularity of this condition is that it is easy to verify the zero-error capacity condition of the channel when the Kraus operator representation is available. In addition, the condition is a lower bound for the zero-error capacity of the channel, that is, $C^{(0)}(\mathcal{E})\geq log\  |N_\mathcal{E}|.$ 
        
        Finally, three examples of quantum channels were presented that possibly have no physical motivation, but mathematically illustrate the proved condition and zero-error capacity. 
        
        About future work, one idea is to use this result to relate the concept of zero-error capacity of quantum channels and Shemesh theorem \cite{Dan}. This idea can bring new results, as well as new relationships between the concept of zero-error capacity and subareas of mathematics such as PI-algebra, i.e., algebras with polynomial identities. 
        
    \section*{Aknowledgements}
        The authors thank CNPq for the financial support.


\begin{thebibliography}{99}
        	
        \bibitem{Cover}  T. M. Cover; J. A. Thomas. \textit{Elements of Information Theory.} John Wiley \& Sons, 2006.
        
        \bibitem{Chuang}  M.A Nielsen; I. L.Chuang\textit{ Quantum Computation and Quantum Information: 10th Anniversary Edition.} [S.l.]: Cambridge University Press, 2010.
        
        \bibitem{Wilde} Mark M. Wilde.\textit{ From Classical to Quantum Shannon Theory}. Cambridge, New York: Cambridge University Press, 2019.
        
        \bibitem{HSW} B. Schumacher and M. D. Westmoreland. Sending classical information via noisy quantum channels. \textit{Phys. Rev. A}, 56(1):131–138, 1997, doi: 10.1103/PhysRevA.56.131.
        
        \bibitem{Holevo} A. S. Holevo. The capacity of the quantum channel with general signal states. \textit{IEEE Trans. Info. Theory}, 44(1):269–273, January 1998, doi: 10.1109/18.651037.
        
        \bibitem{Bennett} C. H. Bennett, P. W. Shor, J. A. Smolin, and A. V. Thapliyal.
        Entanglement-assisted classical capacity of noisy quantum channels.
        \textit{Phys. Rev. Lett.}, 83:3081–3084, May 1999, doi:10.1103/PhysRevLett.83.3081.
        
        \bibitem{Koner} J. Koner and A. Orlitsky. Zero-error information theory. \textit{IEEE Trans.
        Info. Theory,} 44(6):2207–2229, October 1998, doi: 10.1109/18.720537.
        
        \bibitem{Rex0}  Rex A. C. Medeiros and F. M. de Assis. \textit{Quantum Zero-Error Capacity.} International Journal of Quantum Information, vol. 3, n. 1, pp. 135-139, 2005, doi: 10.1142/S0219749905000682.
        
        \bibitem{Shannon} C. E. Shannon. \textit{The zero error capacity of a noisy channel. IRE Trans.} Inform. Theory, IT-2(3):8–19, September  1956, doi: 10.1109/TIT.1956.1056798.
        
        \bibitem{Yamasaki} H. Yamasaki and M. Murao. Quantum State Merging for Arbitrarily
        Small-Dimensional Systems. \textit{IEEE Trans.
        Info. Theory,} 65 (6):3950-3972, June, 2019,  doi: 10.1109/TIT.2018.2889829.
        
        \bibitem{Dereniowski}  D. Dereniowski and M. Jurkiewicz. On the Characteristic Graph of a Discrete Symmetric Channel. \textit{IEEE Trans. Info. Theory,} 67 (6): 3818 - 3823, June, 2021, doi: 10.1109/TIT.2021.3073822.
        
        \bibitem{Rex2} R. A. C. Medeiros, R. Alleaume, G. Cohen and F. M. de Assis. Zero-error capacity of quantum channels and noiseless subsystems.  \textit{International Telecommunications Symposium}, Fortaleza, Brazil, pp. 900-905, 2006, doi: 10.1109/ITS.2006.4433399.
         
        \bibitem{Rex1} Eloá. B. Guedes;  F. M. de Assis; Rex A. C. Medeiros. \textit{Quantum Zero-Error	Information Theory.} Springer, 2016.	
        
        \bibitem{Ved} Ved P. Gupta, Prabha Mandayam e V.S Sunder, \textit{The Functional Analysis of Quantum Information Theory,} Springer, 2015.
        
        \bibitem{Dirac} P. Dirac. \textit{The principles of Quantum Mechanics}, 4th ed. Oxford
        University Press, 1982.
        		
        \bibitem{Dan} D. Shemesh. \textit{Common Elgenvectors of Two Matrices,} Department of Mathematics, Technion - Israel Institute of Technolog, 1984.		
    
    \end{thebibliography}
\end{document}